# Integrating transport equations by combining transfer matrices, transmission-reflection matrices, and Green's matrices, in the context of light propagation in foliage

Antoine Royer
*Département de Génie Physique, Ecole Polytechnique, Montréal, Québec H3C 3A7, Canada*
Sophie Royer
*Institut für Botanik, Universität Innsbruck, Austria*

The propagation of light in a horizontally homogeneous foliage, modeled as a turbid medium, may be expressed in terms of a transport equation of the form

$$\frac{\partial}{\partial z}\begin{pmatrix} \mathbf{D}(z) \\ -\mathbf{U}(z) \end{pmatrix} = \begin{pmatrix} \mathbf{M}_{DD}(z) & \mathbf{M}_{DU}(z) \\ \mathbf{M}_{UD}(z) & \mathbf{M}_{UU}(z) \end{pmatrix} \begin{pmatrix} \mathbf{D}(z) \\ \mathbf{U}(z) \end{pmatrix} + \begin{pmatrix} \mathbf{E}_D(z) \\ \mathbf{E}_U(z) \end{pmatrix}, \qquad \mathbf{U}(z_g) = \mathbf{R}_g \mathbf{D}(z_g) + \mathbf{E}_U(z_g)$$

Here, $z$ is height, $\mathbf{D}(z)$ and $\mathbf{U}(z)$ are vectors of photon fluxes in various downwards ($D$) and upwards ($U$) discrete directions, matrix $\mathbf{M}$ expresses scattering by leaves, $\mathbf{R}_g$ reflection by the ground at $z = z_g$, and $\mathbf{E}(z)$ are emissions. The incident 'down' flux $\mathbf{D}(z_0)$ at the top $z_0$ of the canopy, and emissions, are known. The usual way to get $\mathbf{D}(z), \mathbf{U}(z)$ at all $z$ is to set initially $\mathbf{U}(z) \equiv \mathbf{E}_U(z)$, integrate $\mathbf{D}(z)$ numerically from $z_0$ to $z_g$, then $\mathbf{U}(z)$ from $z_g$ to $z_0$, and iterate. We present a method which is both faster and more accurate, especially if one must do many computations on the same canopy, for different incident fluxes and emissions. This method combines **t**ransfer matrices, **t**ransmission-**r**eflection matrices, and **G**reen's matrices (TTRG). There exist (artificial) extreme light trapping situations for which iterative integration is hardly practical, while TTRG remains as efficient.

## 1. Introduction

In many problems, it is necessary to integrate a transport equation of the form

$$\mathbf{s}\frac{\partial}{\partial z}\mathbf{J}(z) = \mathbf{M}(z)\mathbf{J}(z) + \mathbf{E}(z) \tag{1}$$

where $\mathbf{J}(z)$ and $\mathbf{E}(z)$ are vectors, $\mathbf{s}$ and $\mathbf{M}(z)$ matrices. Typically, the 'scattering matrix' $\mathbf{M}(z)$ and 'source' $\mathbf{E}(z)$ are known, while $\mathbf{J}(z)$, or part of it, is known at some point $z = z_0$, and we want $\mathbf{J}(z)$ at other values of $z$.

The method of numerical integration we present can presumably be adapted to various problems. Here we treat specifically light propagation in a horizontally homogeneous foliage (the 'canopy'), modeled as a turbid medium [1-3]. This has the advantage of being easily visualized, and fairly general in that $\mathbf{M}(z)$ and $\mathbf{E}(z)$ have no special symmetries.

The vector $\mathbf{J}(z)$ of (vertical) photon fluxes (or radiances) separates into two subvectors, $\mathbf{D}(z)$ and $\mathbf{U}(z)$, of 'downward' and 'upward' propagating fluxes. Equation (1) then has the more explicit form, if we denote $\dot{f} \equiv \partial f / \partial z$:

$$\begin{aligned}
\dot{\mathbf{D}}(z) &= \mathbf{E}_D(z) + \mathbf{M}_{DD}(z)\mathbf{D}(z) + \mathbf{M}_{DU}(z)\mathbf{U}(z) & (a) \\
\mathbf{U}(z_g) &= \mathbf{E}_U(z_g) + \mathbf{R}_g \mathbf{D}(z_g) & (b) \\
\dot{\mathbf{U}}(z) &= -\mathbf{E}_U(z) - \mathbf{M}_{UD}(z)\mathbf{D}(z) - \mathbf{M}_{UU}(z)\mathbf{U}(z) & (c)
\end{aligned} \qquad (2)$$

where (b) describes reflection and emission by the ground at $z = z_g$. In practice, only the 'down' flux $\mathbf{D}(z_0)$ at the top $z_0$ of the canopy is known. The usual way to compute $\mathbf{D}(z), \mathbf{U}(z)$ at all heights $z$ is by iteration [1-4]: Set $\mathbf{U}(z) \equiv \mathbf{E}_U(z)$ (or simply $\mathbf{U}(z) \equiv 0$) and compute $\mathbf{D}(z)$ by integrating (a) numerically from $z_0$ to $z_g$; reflect $\mathbf{D}(z_g)$ off the ground with (b); then compute new values of $\mathbf{U}(z)$ by integrating (c) from $z_g$ to $z_0$. Repeat using the new values of $\mathbf{U}(z)$, and so on until $\mathbf{D}(z)$ and $\mathbf{U}(z)$ stabilize.

Each iteration in effect allows photons to undergo additional scatterings. For visible and thermal radiation, which are strongly absorbed by leaves (above 80%), photons have a high probability of getting absorbed at each scattering, so that convergence is fast (4 or 5 iterations suffice); for near-infra-red light (NIR), absorption is less than 20%, and more iterations are required. One can devise (artificial) extreme light trapping situations requiring any number of iterations (hundreds of thousands), and for which, moreover, specifying a convergence criterium is problematic, so that iterative integration is hardly practical. Monte Carlo calculations [5], whereby histories of many individual photons are computed, fare even worse. But the method presented in this paper remains as fast and accurate.

This method combines transfer matrices, transmission-reflection matrices, and Green's matrices (TTRG). All these matrices are specific to the canopy, hence need to be computed only once. They are then applied to vectors comprised of incident and emitted radiances, to quickly yield the light climate (the set of fluxes everywhere). This is especially advantageous if one has to do many computations on the same canopy, for different incident radiances and/or leaf thermal emissions. Indeed, the latter depend non-linearly on leaf temperatures, which themselves depend non-linearly on leaf absorption rates, so that the only feasible way to determine leaf temperatures is by iteration. Whence repeated computations of the thermal light climate.

Transfer matrices can be used *alone* for moderately thick canopies, but run into numerical problems with thick canopies. Aronson and coworkers [6] formulated a transfer matrix method applicable to optically thick media, but for cases that $\mathbf{M}(z)$ is constant over sizable $z$ intervals, and the symmetries $\mathbf{M}_{DU} = \mathbf{M}_{UD}$, $\mathbf{M}_{UU} = \mathbf{M}_{DD}$ hold. These conditions are not met in our problem, as leaf densities vary with $z$, and the top and bottom of leaves have different optical properties. Transmission-reflection matrices [7], as well as Green's matrices, can also be used alone, but are not optimal, as will be described onwards. It is more efficient to suitably combine the use of these different types of matrices.

The paper has three parts. Part I establishes the light transport equation (LTE). Part II presents TTRG. Part III describes other methods of numerically integrating the LTE, for comparison with TTRG. The Conclusion briefly presents results of numerical tests, to be described in detail elsewhere. Future papers will apply TTRG to real canopies, including calculations of leaf temperatures and rates of photosynthesis.

**Part I   Starting equations**

We start with a general discussion of the problem (section 2), and then establish the light transport equation (section 3).

**2   General aspects**

The foliage is modeled as a turbid medium of infinitesimal leaf elements, described by volume densities of leaf areas of various kinds and orientations [1-4]. Photons are scattered, absorbed or emitted by leaves and the ground below. We wish to calculate the light climate inside the canopy, namely the set of photon fluxes $J(\Omega,\mathbf{r})$ in all directions $\Omega$, at every point $\mathbf{r} = (x,y,z)$, resulting from the penetration of sun and sky light into the canopy, and from thermal emissions by leaves and ground.

Consider a *narrow* beam of light entering the canopy in direction $\Omega_i$. When a photon hits a leaf, it is either absorbed, or scattered into any one of a continuum of directions $\Omega_f$. So when the light beam, comprised of myriad photons, strikes a leaf, part is absorbed, and the rest *fans out* into a continuum of scattered beams. All these beams likewise fan out as they in turn hit other leaves, and so on. On this account, calculating $J(\Omega,\mathbf{r})$ seems a formidable affair.

One way to tackle the problem is to follow *individual* photons rather than light beams. Each scattering now has a *single* random outcome, so that a given photon can have many different histories. By calculating and combining many such histories, one reproduces the multiple fanning out of a beam. Doing this for many different incident directions $\Omega_i$ and points of entry into the canopy, and finally summing, one obtains $J(\Omega,\mathbf{r})$ to any accuracy, in principle. Such 'Monte-Carlo' calculations [5] are of course very time consuming.

Another approach is to break up $J(\Omega,\mathbf{r})$ into subclimates $J^{(k)}(\Omega,\mathbf{r})$, $k = 0,1,2,...$ comprising photons that reached $\mathbf{r}$ after scattering $k$ times off leaves or ground. Then,

$$J(\Omega,\mathbf{r}) = J^{(0)}(\Omega,\mathbf{r}) + J^{(1)}(\Omega,\mathbf{r}) + J^{(2)}(\Omega,\mathbf{r}) + ... \tag{3}$$

The $J^{(k)}$ can be calculated recursively in a straightforward manner [8]. Convergence of the *multiple-scattering expansion* (3) is rapid if absorption is strong (visible and thermal radiation), less so otherwise (near-infrared), and hopelessly slow in cases of extreme light trapping.

Still another approach, and the one we use, is to deal directly with the whole light climate $J(\Omega,\mathbf{r})$, and obtain for it an integro-differential equation called the *light transport equation* (LTE) [9]. That it is possible to do so is due to the fact that over an infinitesimal distance $d\ell$, a photon can hit *at most one* leaf element (to first order in $d\ell$), so that the photon flux is affected in a *simple* way. Integrating the LTE iteratively in fact reproduces the series (3).

## 3  The light transport equation (LTE)

We let the positive $z$ direction point *downwards*. We deal with a set of discrete directions $\Omega_j = (\theta_j, \varphi_j)$, $j = 1, 2, ..., N_J$, and write $j \in D$ or $j \in U$ according as $\Omega_j$ points downwards or upwards. Vectors $\mathbf{J}$, and matrices $\mathbf{M}$, will often be written in 'down-up' block form:

$$\mathbf{J} = \begin{pmatrix} \mathbf{J}_D \\ \mathbf{J}_U \end{pmatrix}, \qquad \mathbf{M} = \begin{pmatrix} \mathbf{M}_{DD} & \mathbf{M}_{DU} \\ \mathbf{M}_{UD} & \mathbf{M}_{UU} \end{pmatrix} \tag{4}$$

where $\mathbf{J}_D$ has elements $J_j$, $j \in D$, and $\mathbf{M}_{DU}$ has elements $M_{ij}$, $i \in D$, $j \in U$, etc.

Assuming horizontally homogeneous conditions, we denote by $J_j(z)$ the *vertical* flux of photons $j$ (i.e., travelling in direction $\Omega_j$), so that $J_j(z)$ is the number of photons $j$ crossing per second a unit horizontal area at height $z$. Consider now a horizontal layer $(z_1, z_2)$ of infinitesimal thickness $dz = z_2 - z_1 > 0$. Let $g_i$ be the probability for a photon $i$ to hit a leaf within the layer, and $h_{fi}$ that to get scattered into direction $f$. Of course, the probability $\sum_f h_{fi}$ to get scattered into *any* direction is less than that to get intercepted, $g_i$. The difference

$$a_i = g_i - \sum_f h_{fi} \geq 0 \tag{5}$$

is the probability of absorbtion. Finally, let $e_f$ be the number of photons emitted per second in direction $f$ by a unit area of the layer. Across the layer, the vertical flux $J_f(z)$ looses photons due to interception by leaves, but gains photons scattered or emitted in direction $f$. Thus, denoting $\mathbf{J}(z_n) = \mathbf{J}_n = (\mathbf{D}_n, \mathbf{U}_n)$, $n = 1, 2$:

$$\begin{aligned} \mathbf{D}_2 &= (1 - \mathbf{g}_D)\mathbf{D}_1 + \mathbf{h}_{DD}\mathbf{D}_1 + \mathbf{h}_{DU}\mathbf{U}_2 + \mathbf{e}_D \\ \mathbf{U}_1 &= (1 - \mathbf{g}_U)\mathbf{U}_2 + \mathbf{h}_{UU}\mathbf{U}_2 + \mathbf{h}_{UD}\mathbf{D}_1 + \mathbf{e}_U \end{aligned} \quad (6)$$

where $g_{ij} = g_j \delta_{ij}$. Clearly, $\mathbf{g}, \mathbf{h}, \mathbf{e}$ are infinitesimals proportional to $dz$:

$$\mathbf{g} = \mathbf{G}dz, \quad \mathbf{h} = \mathbf{H}dz, \quad \mathbf{e} = \mathbf{E}dz \quad (dz = z_2 - z_1 > 0) \quad (7)$$

where $\mathbf{G}, \mathbf{H}, \mathbf{E}$ are probabilities (or rates) per unit vertical distance. These are readily calculated in terms of the volume densities of leaf areas (of various species and orientations), and of the scattering coefficients and emission rates for individual leaves [1-4]. We may rewrite (6) as

$$\mathbf{J}^{out} \equiv \begin{pmatrix} \mathbf{D}_2 \\ \mathbf{U}_1 \end{pmatrix} = (\mathbf{1} + \mathbf{M}dz)\mathbf{J}^{in} + \mathbf{E}dz, \quad \mathbf{M} \equiv \mathbf{H} - \mathbf{G}, \quad \mathbf{J}^{in} \equiv \begin{pmatrix} \mathbf{D}_1 \\ \mathbf{U}_2 \end{pmatrix} \quad (8)$$

Since $(\mathbf{1} - \mathbf{g}_U)^{-1} \approx \mathbf{1} + \mathbf{g}_U$ and $\mathbf{g}_U \mathbf{U}_2 \approx \mathbf{g}_U \mathbf{U}_1$ to first order in $dz$, (6) may also be written as

$$\begin{aligned} \mathbf{D}_2 &= (\mathbf{1} - \mathbf{g}_D)\mathbf{D}_1 + \mathbf{h}_{DD}\mathbf{D}_1 + \mathbf{h}_{DU}\mathbf{U}_1 + \mathbf{e}_D \\ \mathbf{U}_2 &= (\mathbf{1} + \mathbf{g}_U)\mathbf{U}_1 - \mathbf{h}_{UU}\mathbf{U}_1 - \mathbf{h}_{UD}\mathbf{D}_1 - \mathbf{e}_U \end{aligned} \quad (9)$$

Note that (6),(8) relate the fluxes $\mathbf{J}^{out}$ exiting the layer $(z_1, z_2)$ to $\mathbf{J}^{in}$ entering it, while (9) relates $\mathbf{J}_2 = (\mathbf{D}_2, \mathbf{U}_2)$ just below the layer to $\mathbf{J}_1 = (\mathbf{D}_1, \mathbf{U}_1)$ just above. Subtituting (7) into (9), and using $\mathbf{J}_2 - \mathbf{J}_1 = (\partial \mathbf{J}/\partial z)dz$, we recover equation (1):

$$\mathbf{s}\frac{\partial}{\partial z}\mathbf{J}(z) = \mathbf{M}(z)\mathbf{J}(z) + \mathbf{E}(z), \quad \mathbf{M}(z) = -\mathbf{G}(z) + \mathbf{H}(z), \quad \mathbf{s} \equiv \begin{pmatrix} 1 & 0 \\ 0 & -1 \end{pmatrix} \quad (10)$$

**Part II  The TTRG method**

We start by describing the transfer matrix method (TMM), and analyze why it fails when applied numerically to thick canopies. Our remedy is to partition the canopy into moderately thick 'medium' layers, and apply TMM recursively. This amounts to a discrete version of the 'invariant embedding method' [10], but like the latter, it only yields the light reflected and emitted by the canopy. To get the light climate inside the canopy, we first obtain the radiances between 'medium' layers by using transmission-reflection matrices and a Green's matrix;

radiances *within* 'medium' layers can then be obtained using transfer matrices.

## 4 Transfer matrix method

Transfer matrices $\mathbf{T}(z,z')$ are defined by

$$\frac{\partial}{\partial z}\mathbf{T}(z,z') = \mathbf{sM}(z)\mathbf{T}(z,z'), \qquad \mathbf{T}(z,z) = \mathbf{1}, \qquad \mathbf{T}(z,z')\mathbf{T}(z',z'') = \mathbf{T}(z,z'') \qquad (11)$$

If $\mathbf{M}(z) \equiv \mathbf{M}$ is constant between $z$ and $z'$, then $\mathbf{T}(z,z') = e^{(z-z')\mathbf{sM}}$. So let us partition the canopy $(z_0, z_g)$ into $N$ 'thin' layers $\Delta z_n$:

$$(z_0, z_g) = (z_0, z_1, ..., z_N), \qquad \Delta z_n = z_n - z_{n-1} \qquad (n = 1, 2, ..., N) \qquad (12)$$

with $z_0$ at the top of the canopy, $z_N = z_g$ at the bottom. These 'thin' layers are chosen thin enough that $\mathbf{M}(z) \approx \mathbf{M}_n$ and $\mathbf{E}(z) \approx \mathbf{E}_n$ if $z \in (z_{n-1}, z_n)$, and also that $|M_{n,ij}\Delta z_n| \ll 1$ for all $n$, $i, j$. In this way, the thin layer transfer matrices

$$\mathbf{T}_n \equiv \mathbf{T}(z_n, z_{n-1}) \approx e^{\mathbf{sM}_n \Delta z_n} = \mathbf{1} + \mathbf{sM}_n \Delta z_n + \tfrac{1}{2}(\mathbf{sM}_n)^2 \Delta z_n^2 + ... \qquad (13)$$

differ little from $\mathbf{1}$. We have $\mathbf{T}(z_n, z_0) = \mathbf{T}_n \mathbf{T}_{n-1} \cdots \mathbf{T}_1$.

Assuming that $\mathbf{J}_0 \equiv \mathbf{J}(z_0)$ is known, the solution of (10) may be written as

$$\mathbf{J}(z) = \mathbf{T}(z,z_0)\mathbf{J}(z_0) + \mathbf{f}(z,z_0), \qquad \mathbf{f}(z,z_0) \equiv \int_{z_0}^{z} dz' \, \mathbf{T}(z,z')\mathbf{sE}(z') \qquad (14)(a,b)$$

If $\mathbf{E}(z) \equiv 0$, then $\mathbf{J}(z) = \mathbf{T}(z,z_0)\mathbf{J}(z_0)$, showing that $\mathbf{T}(z,z_0)$ 'propagates' $\mathbf{J}(z_0)$ from $z_0$ to $z$. We call $\mathbf{f}(z,z_0)$ 'propagated emissions', since $\mathbf{T}(z,z')\mathbf{sE}(z')$ is the emission at $z'$ propagated to $z$. In numerical computations, we only want $\mathbf{J}(z_n)$ at the discrete heights $z_n$. These are best obtained recursively, i.e., approximating $\mathbf{f}(z_n, z_{n-1}) \approx \mathbf{sE}_n \Delta z_n$:

$$\mathbf{J}_n = \mathbf{T}_n \mathbf{J}_{n-1} + \mathbf{se}_n, \qquad \mathbf{e}_n = \mathbf{E}_n \Delta z_n, \qquad \mathbf{J}_n \equiv \mathbf{J}(z_n), \qquad (n = 1, 2, ..., N) \qquad (15)$$

In practice, only the 'down' component $\mathbf{D}_0 = \mathbf{D}(z_0)$ of $\mathbf{J}_0$ is known, namely the sunlight and skylight incident on the top $z = z_0$ of the canopy. The 'up' component $\mathbf{U}_0 = \mathbf{U}(z_0)$ (the light reflected and emitted by the canopy) is unknown.

***Determination of*** $\mathbf{U}_0$: Let $\mathbf{T}$ and $\mathbf{f}$ pertain to the whole canopy $(z_0, z_g)$:

$$\mathbf{T} \equiv \mathbf{T}(z_g, z_0), \qquad \mathbf{f} \equiv \mathbf{f}(z_g, z_0) \qquad (16)$$

Putting $\mathbf{J}(z_0) = \mathbf{J}_0 = (\mathbf{D}_0, \mathbf{U}_0)$, etc., write (14) for $z = z_g$ as $\mathbf{J}_g = \mathbf{T}\mathbf{J}_0 + \mathbf{f}$, that is:

$$\begin{pmatrix} \mathbf{D}_g \\ \mathbf{U}_g \end{pmatrix} = \begin{pmatrix} \mathbf{T}_{DD} & \mathbf{T}_{DU} \\ \mathbf{T}_{UD} & \mathbf{T}_{UU} \end{pmatrix} \begin{pmatrix} \mathbf{D}_0 \\ \mathbf{U}_0 \end{pmatrix} + \begin{pmatrix} \mathbf{f}_D \\ \mathbf{f}_U \end{pmatrix} = \begin{pmatrix} \mathbf{T}_{DD}\mathbf{D}_0 + \mathbf{T}_{DU}\mathbf{U}_0 + \mathbf{f}_D \\ \mathbf{T}_{UD}\mathbf{D}_0 + \mathbf{T}_{UU}\mathbf{U}_0 + \mathbf{f}_U \end{pmatrix} \tag{17}$$

Of the four fluxes $\mathbf{D}_0, \mathbf{U}_0, \mathbf{D}_g, \mathbf{U}_g$, only $\mathbf{D}_0$ is known. However, we also have at $z_g$:

$$\mathbf{U}_g = \mathbf{R}_g \mathbf{D}_g + \mathbf{e}_g, \qquad \mathbf{e}_g \equiv \mathbf{E}_U(z_g) \tag{18}$$

Substituting $\mathbf{D}_g, \mathbf{U}_g$, as given by (17), into (18), and then solving for $\mathbf{U}_0$, we get:

$$\mathbf{U}_0 = \mathbf{R}_0 \mathbf{D}_0 + \mathbf{e}_0 \tag{19}$$

$$\mathbf{R}_0 = \frac{1}{\mathbf{T}_{UU} - \mathbf{R}_g \mathbf{T}_{DU}} \left( \mathbf{R}_g \mathbf{T}_{DD} - \mathbf{T}_{UD} \right), \qquad \mathbf{e}_0 = \frac{1}{\mathbf{T}_{UU} - \mathbf{R}_g \mathbf{T}_{DU}} \left( \mathbf{R}_g \mathbf{f}_D - \mathbf{f}_U + \mathbf{e}_g \right) \tag{20}$$

Comparing (19) with (18), we see that $\mathbf{e}_0$ is an 'up' emission vector, and $\mathbf{R}_0$ a 'down-to-up' reflection matrix, from the surface $z = z_0$ (thus, $e_{0,j} \geq 0$ and $0 \leq R_{0,ij} \leq 1$).

Once $\mathbf{U}_0$, hence the complete $\mathbf{J}_0$, are known, the fluxes $\mathbf{J}_n$ between thin layers can be computed recursively using (15). However, in numerical computations on thick canopies, unphysical negative elements appear in $\mathbf{R}_0$ and $\mathbf{e}_0$, hence in $\mathbf{U}_0$ (negative fluxes). To trace the origin of this nonsense, let us first define transmission-reflection matrices.

## 5 Transmission-reflection matrix, and canopy emission vector

Equation (17) relates $\mathbf{J}_g = (\mathbf{D}_g, \mathbf{U}_g)$ *below* the canopy to $\mathbf{J}_0 = (\mathbf{D}_0, \mathbf{U}_0)$ *above*. Let us rather relate $\mathbf{J}^{out} = (\mathbf{D}_g, \mathbf{U}_0)$ coming *out* of the canopy, to $\mathbf{J}^{in} = (\mathbf{D}_0, \mathbf{U}_g)$ coming *in*. From (17) we deduce $\mathbf{U}_0 = \mathbf{T}_{UU}^{-1}(\mathbf{U}_g - \mathbf{T}_{UD}\mathbf{D}_0 - \mathbf{f}_U)$, where $\mathbf{T}_{UU}^{-1} \equiv (\mathbf{T}_{UU})^{-1}$. We can then write:

$$\mathbf{J}^{out} \equiv \begin{pmatrix} \mathbf{D}_g \\ \mathbf{U}_0 \end{pmatrix} = \begin{pmatrix} \mathbf{t}\mathbf{D}_0 + \mathbf{r}\mathbf{U}_g + \mathbf{d} \\ \mathbf{\rho}\mathbf{D}_0 + \mathbf{\tau}\mathbf{U}_g + \mathbf{u} \end{pmatrix} = \mathbf{K}\mathbf{J}^{in} + \mathbf{\varepsilon}, \qquad \mathbf{J}^{in} \equiv \begin{pmatrix} \mathbf{D}_0 \\ \mathbf{U}_g \end{pmatrix} \tag{21}$$

where we define the *transmission-reflection* matrix $\mathbf{K}$, and *canopy emission vector* $\mathbf{\varepsilon}$, by

$$\mathbf{K} \equiv \begin{pmatrix} \mathbf{t} & \mathbf{r} \\ \mathbf{\rho} & \mathbf{\tau} \end{pmatrix} \equiv \begin{pmatrix} \mathbf{T}_{DD} - \mathbf{T}_{DU}\mathbf{T}_{UU}^{-1}\mathbf{T}_{UD} & \mathbf{T}_{DU}\mathbf{T}_{UU}^{-1} \\ -\mathbf{T}_{UU}^{-1}\mathbf{T}_{UD} & \mathbf{T}_{UU}^{-1} \end{pmatrix}, \qquad \mathbf{\varepsilon} \equiv \begin{pmatrix} \mathbf{d} \\ \mathbf{u} \end{pmatrix} \equiv \begin{pmatrix} \mathbf{f}_D - \mathbf{r}\mathbf{f}_U \\ -\mathbf{\tau}\mathbf{f}_U \end{pmatrix} \tag{22}$$

Considering $\mathbf{D}_g = \mathbf{t}\mathbf{D}_0 + \mathbf{r}\mathbf{U}_g + \mathbf{d}$, we see that $\mathbf{t}$ is 'down' transmission through the canopy, $\mathbf{r}$ is 'up-to-down' reflection off the surface $z = z_g$, and $\mathbf{d}$ is 'down' emission from $z_g$. Similarly for $\mathbf{\tau}, \mathbf{\rho}$, and $\mathbf{u}$. Their physical meanings imply that $\varepsilon_j \geq 0$ and $0 \leq K_{ij} \leq 1$. Note that $\mathbf{K}, \mathbf{\varepsilon}$ pertain to the canopy *alone*, and $\mathbf{\rho} = -\mathbf{T}_{UU}^{-1}\mathbf{T}_{UD}$ and $\mathbf{u} = -\mathbf{T}_{UU}^{-1}\mathbf{f}_U$ are identical to $\mathbf{R}_0$ and $\mathbf{e}_0$ in

(20), if $\mathbf{R}_g = \mathbf{e}_g = 0$ (as is clear physically). Reciprocally to (22):

$$\mathbf{T} = \begin{pmatrix} \mathbf{T}_{DD} & \mathbf{T}_{DU} \\ \mathbf{T}_{UD} & \mathbf{T}_{UU} \end{pmatrix} = \begin{pmatrix} \mathbf{t} - \mathbf{r}\boldsymbol{\tau}^{-1}\boldsymbol{\rho} & \mathbf{r}\boldsymbol{\tau}^{-1} \\ -\boldsymbol{\tau}^{-1}\boldsymbol{\rho} & \boldsymbol{\tau}^{-1} \end{pmatrix}, \qquad \mathbf{f} = \begin{pmatrix} \mathbf{f}_D \\ \mathbf{f}_U \end{pmatrix} = \begin{pmatrix} \mathbf{d} - \mathbf{r}\boldsymbol{\tau}^{-1}\mathbf{u} \\ -\boldsymbol{\tau}^{-1}\mathbf{u} \end{pmatrix} \qquad (23)$$

Indeed, if there are no emissions and $\mathbf{D}_0 = 0$, then $\mathbf{U}_g = \mathbf{T}_{UU}\mathbf{U}_0 = \mathbf{T}_{UU}\boldsymbol{\tau}\mathbf{U}_g$, so that $\mathbf{T}_{UU}\boldsymbol{\tau} = \mathbf{1}$ (since $\mathbf{U}_g$ can be any non-negative vector). Hence, since $\mathbf{T}_{UU}$ and $\boldsymbol{\tau}$ necessarily exist, so do $\mathbf{T}_{UU}^{-1}$ and $\boldsymbol{\tau}^{-1}$. Also, in (20), $\mathbf{T}_{UU} - \mathbf{R}_g\mathbf{T}_{DU} = (\mathbf{1} - \mathbf{R}_g\mathbf{r})\boldsymbol{\tau}^{-1}$, by (23), is invertible. Indeed, the series $\mathbf{1} + \mathbf{R}_g\mathbf{r} + \mathbf{R}_g\mathbf{r}\mathbf{R}_g\mathbf{r} + ...$ must converge, as it sums over all back and forth reflections between ground and canopy, and a *finite* canopy cannot be perfectly reflecting, so that each reflection cycle entails some loss (hence no infinite buildup of radiance).

## 6  Numerical limitations of the transfer matrix method

It is useful to define, for any layer $(z, z')$ of canopy, its *leaf area index*:

$$\text{LAI} = \textit{total leaf area inside the layer, per unit horizontal area of canopy} \qquad (24)$$

The transfer matrix method (section 4) is quick and simple in principle. But as already said, in numerical computations on thick canopies (LAI above $\sim 4$), there appear negative elements in $\mathbf{R}_0$ and $\mathbf{e}_0$. We can now see why: A realistic thick canopy reflects about 5% of the incident light, and transmits almost none. This means that the reflection matrices $\mathbf{r}$ and $\boldsymbol{\rho}$ are fairly small, and the transmission matrices $\mathbf{t}$ and $\boldsymbol{\tau}$ *very* small[1]. It follows that $\boldsymbol{\tau}^{-1}$ is very large, whence also all four blocks of $\mathbf{T}$ in (23). Suppose now zero emissions, and $\mathbf{U}_g = 0$, so that $\mathbf{U}_0 = \boldsymbol{\rho}\mathbf{D}_0$ and $\mathbf{D}_g = \mathbf{t}\mathbf{D}_0$. However, the transfer matrix expression

$$\mathbf{D}_g = \mathbf{T}_{DD}\mathbf{D}_0 + \mathbf{T}_{DU}\mathbf{U}_0 = (\mathbf{t} - \mathbf{r}\boldsymbol{\tau}^{-1}\boldsymbol{\rho})\mathbf{D}_0 + (\mathbf{r}\boldsymbol{\tau}^{-1})\boldsymbol{\rho}\mathbf{D}_0 \qquad (25)$$

expresses the *very* small flux $\mathbf{D}_g = \mathbf{t}\mathbf{D}_0$ as the *difference* of two *very* large fluxes. This evidently yields nonsense if $\mathbf{T}$ is larger than $10^{n_c}$, where $n_c$ is the number of digits carried by the computer ($n_c = 15$ in double precision). Similarly, $\mathbf{R}_0, \mathbf{e}_0$ in (20) involve $\mathbf{T}_{UU} - \mathbf{R}_g\mathbf{T}_{DU}$ and $\mathbf{R}_g\mathbf{T}_{DD} - \mathbf{T}_{UD}$, both differences of very large matrices. Clearly then, the simple transfer matrix method is applicable only to moderately thick canopies.

---

[1] The 'size' of a matrix or vector is here understood as the size of its *largest* element.

We will give in section 7 a way of obtaining (non-negative) $\mathbf{R}_0, \mathbf{e}_0$ for thick canopies. Meanwhile, we warn that even with $\mathbf{U}_0$ known exactly, other problems arise: For when the $\mathbf{J}_n$ between thin layers are computed recursively using (15), negative values show up in $\mathbf{U}_n$ below a certain depth (below a cumulative LAI of about 5), even though the numerical pitfall (25) is avoided since (15) builds up $\mathbf{D}_g$ step by step. To see what happens, observe that for 'thin' layers, $\mathbf{T}_n = \mathbf{T}(z_n, z_{n-1}) \approx \mathbf{1} + \mathbf{M}_n \Delta z$, by (13), so that equations (15) are of the form (9):

$$\begin{aligned}
\mathbf{D}_n &\approx (\mathbf{1} - \mathbf{g}_D)\mathbf{D}_{n-1} + \mathbf{h}_{DD}\mathbf{D}_{n-1} + \mathbf{h}_{DU}\mathbf{U}_{n-1} + \mathbf{e}_D & (a) \\
\mathbf{U}_n &\approx (\mathbf{1} + \mathbf{g}_U)\mathbf{U}_{n-1} - \mathbf{h}_{UU}\mathbf{U}_{n-1} - \mathbf{h}_{UD}\mathbf{D}_{n-1} - \mathbf{e}_U & (b)
\end{aligned} \quad (26)$$

where $\mathbf{g}, \mathbf{h}, \mathbf{e}$ are much less than one. We now notice that whereas $\mathbf{D}_n$ gets attenuated (as $n$ increases) via *multiplication* by numbers less than 1 (all coefficients in (26)(a) are non-negative, and $\leq 1$), $\mathbf{U}_n$ must approach zero by way of *substractions*, since the elements of the (diagonal) matrix $\mathbf{1} + \mathbf{g}_U$ are $\geq 1$. Whence the likelihood of undershooting and going slightly negative at some point $z_c$ (and thenceforth more and more negative). One might simply set to zero all radiances below $z_c$, since the exact fluxes are anyhow very small there. But this will not do if we wish to assess the small but extant photosynthesis in the undergrowth.

## 7 Multistep calculation of $\mathbf{R}_0$ and $\mathbf{e}_0$ for thick canopies

Partition the canopy into $M$ 'medium' layers $(z_{m-1}, z_m)$, $m = 1, 2, ..., M$, with $z_0$ at the top, $z_M = z_g$ at the bottom. These 'medium' layers are chosen (LAI less than ~3) such that their transfer matrices have no overly large elements. Denote, for each 'medium' layer $m$:

$$\mathbf{T}^{(m)} \equiv \mathbf{T}(z_m, z_{m-1}) = \begin{pmatrix} \mathbf{T}^{(m)}_{DD} & \mathbf{T}^{(m)}_{DU} \\ \mathbf{T}^{(m)}_{UD} & \mathbf{T}^{(m)}_{UU} \end{pmatrix}, \qquad \mathbf{f}^{(m)} \equiv \mathbf{f}(z_m, z_{m-1}) = \begin{pmatrix} \mathbf{f}^{(m)}_D \\ \mathbf{f}^{(m)}_U \end{pmatrix} \quad (27)$$

Also, let $\mathbf{R}_m$ be the 'down-to-up' reflection matrix, and $\mathbf{e}_m$ the 'up' emission vector, from the surface $z = z_m$ (similar to $\mathbf{R}_0$ and $\mathbf{e}_0$, but for the slab $(z_m, z_g)$ plus ground). Since $\mathbf{R}_M = \mathbf{R}_g$ and $\mathbf{e}_M = \mathbf{e}_g$ are known, $(\mathbf{R}_{M-1}, \mathbf{e}_{M-1}), ..., (\mathbf{R}_0, \mathbf{e}_0)$ can be computed recursively by using

$$\begin{aligned}
\mathbf{R}_{m-1} &= \frac{1}{\mathbf{T}^{(m)}_{UU} - \mathbf{R}_m \mathbf{T}^{(m)}_{DU}} \left( \mathbf{R}_m \mathbf{T}^{(m)}_{DD} - \mathbf{T}^{(m)}_{UD} \right) \\
\mathbf{e}_{m-1} &= \frac{1}{\mathbf{T}^{(m)}_{UU} - \mathbf{R}_m \mathbf{T}^{(m)}_{DU}} \left( \mathbf{R}_m \mathbf{f}^{(m)}_D - \mathbf{f}^{(m)}_U + \mathbf{e}_m \right)
\end{aligned} \quad (m = M, M-1, ..., 1) \quad (28)$$

wherein $\mathbf{R}_m$ and $\mathbf{e}_m$ play the role of $\mathbf{R}_g$ and $\mathbf{e}_g$ in (20). Since $\mathbf{R}_M = \mathbf{R}_g$ and $\mathbf{e}_M = \mathbf{e}_g$ are non-negative, so will $\mathbf{R}_{M-1}, \mathbf{e}_{M-1}$, in practice as well as in principle, since $\mathbf{T}^{(M)}$ is not overly large; whence also $\mathbf{R}_{M-2}, \mathbf{e}_{M-2}$, and so on until $\mathbf{R}_0, \mathbf{e}_0$. This multistep calculation of $\mathbf{R}_0, \mathbf{e}_0$ amounts to a discrete version of the *invariant embedding method* [10] (see section 16).

***Discussion***: If one is only interested in the 'up' light $\mathbf{U}_0 = \mathbf{R}_0 \mathbf{D}_0 + \mathbf{e}_0$ reflected and emitted by the canopy (as in remote sensing), then the above suffices. But if we also need the photon fluxes $\mathbf{J}(z)$ inside the canopy (e.g., to compute rates of photosynthesis), then we want $\mathbf{J}_n$ between thin layers. But as explained in section 6, even with $\mathbf{U}_0$ known exactly, negative fluxes $\mathbf{U}_n$ may show up if $\mathbf{J}_n$ are computed recursively from $z_0$ to $z_g$ using (15).

Our remedy will be to first get $\mathbf{J}_m \equiv \mathbf{J}(z_m)$ between *medium* layers, including $\mathbf{U}_0$ (thus $\mathbf{R}_0, \mathbf{e}_0$ are not needed, so that (28) is useful only if one only wants $\mathbf{U}_0$). The $\mathbf{J}_n$ between *thin* layers *within* each medium layer $(z_{m-1}, z_m)$ can then be computed recursively from $z_{m-1}$ to $z_m$ by using (15) (which is no problem over moderately thick layers).

## 8 Fluxes between 'medium' layers, Green's matrix

Let $\mathbf{K}_m, \boldsymbol{\varepsilon}_m$ pertain to 'medium' layer $m$, so that similarly to (21):

$$\mathbf{J}_m^{out} \equiv \begin{pmatrix} \mathbf{D}_m \\ \mathbf{U}_{m-1} \end{pmatrix} = \boldsymbol{\varepsilon}_m + \mathbf{K}_m \mathbf{J}_m^{in} = \begin{pmatrix} \mathbf{d}_m \\ \mathbf{u}_m \end{pmatrix} + \begin{pmatrix} \mathbf{t}_m & \mathbf{r}_m \\ \boldsymbol{\rho}_m & \boldsymbol{\tau}_m \end{pmatrix} \begin{pmatrix} \mathbf{D}_{m-1} \\ \mathbf{U}_m \end{pmatrix}, \qquad \mathbf{J}_m^{in} \equiv \begin{pmatrix} \mathbf{D}_{m-1} \\ \mathbf{U}_m \end{pmatrix} \qquad (29)$$

where $\mathbf{J}_m \equiv \mathbf{J}(z_m)$. We thus have, adding in the ground equation at $z_M = z_g$:

$$\begin{aligned} \mathbf{D}_m &= \mathbf{d}_m + \mathbf{t}_m \mathbf{D}_{m-1} + \mathbf{r}_m \mathbf{U}_m & (m = 1, 2, ..., M) & \quad (a) \\ \mathbf{U}_{m-1} &= \mathbf{u}_m + \boldsymbol{\rho}_m \mathbf{D}_{m-1} + \boldsymbol{\tau}_m \mathbf{U}_m & (m = 1, 2, ..., M+1) & \quad (b) \end{aligned} \qquad (30)$$

$$\mathbf{u}_{M+1} \equiv \mathbf{e}_g, \qquad \boldsymbol{\rho}_{M+1} \equiv \mathbf{R}_g, \qquad \boldsymbol{\tau}_{M+1} \equiv 0 \qquad (31)$$

where (b) with $m = M+1$ is the ground equation $\mathbf{U}_M = \mathbf{e}_g + \mathbf{R}_g \mathbf{D}_M$.

***Green's matrix***: Remember that $\mathbf{J}_m = (\mathbf{D}_m, \mathbf{U}_m)$ is a vector of dimension $N_J = N_D + N_U$, with elements $(\mathbf{J}_m)_j$. Let us view $J_{(m,j)} \equiv (\mathbf{J}_m)_j$ as elements of a large vector $J$ of dimension $N_G = M \times N_J$, with the pair $(m, j)$ acting as a single vector index. Then, viewing the incident light $\mathbf{D}_0$ as an 'emission' at $z_0$, we may write equations (30) as

$$J = \mathcal{E} + \mathcal{Q} J, \qquad \mathbf{U}_0 = \mathbf{u}_1 + \boldsymbol{\rho}_1 \mathbf{D}_0 + \boldsymbol{\tau}_1 \mathbf{U}_1 \qquad (32)$$

where the light climate vector $J$, 'emission' vector $\mathcal{E}$, and matrix $\mathcal{Q}$, are given by (we display the case of three 'medium' layers, $M = 3$):

$$J = \begin{pmatrix} J_D \\ J_U \end{pmatrix}, \qquad \mathcal{E} = \begin{pmatrix} \mathcal{E}_D \\ \mathcal{E}_U \end{pmatrix}, \qquad \mathcal{Q} = \begin{pmatrix} \mathcal{T}_d & \mathcal{R}_\wedge \\ \mathcal{R}_\vee & \mathcal{T}_u \end{pmatrix}$$

$$J_D = \begin{pmatrix} \mathbf{D}_1 \\ \mathbf{D}_2 \\ \mathbf{D}_3 \end{pmatrix}, \quad J_U = \begin{pmatrix} \mathbf{U}_1 \\ \mathbf{U}_2 \\ \mathbf{U}_3 \end{pmatrix}, \quad \mathcal{E}_D = \begin{pmatrix} \mathbf{d}_1 + \mathbf{t}_1 \mathbf{D}_0 \\ \mathbf{d}_2 \\ \mathbf{d}_3 \end{pmatrix}, \quad \mathcal{E}_U = \begin{pmatrix} \mathbf{u}_2 \\ \mathbf{u}_3 \\ \mathbf{e}_g \end{pmatrix} \qquad (33)$$

$$\mathcal{T}_d = \begin{pmatrix} 0 & 0 & 0 \\ \mathbf{t}_2 & 0 & 0 \\ 0 & \mathbf{t}_3 & 0 \end{pmatrix}, \; \mathcal{T}_u = \begin{pmatrix} 0 & \mathbf{\tau}_2 & 0 \\ 0 & 0 & \mathbf{\tau}_3 \\ 0 & 0 & 0 \end{pmatrix}, \; \mathcal{R}_\wedge = \begin{pmatrix} \mathbf{r}_1 & 0 & 0 \\ 0 & \mathbf{r}_2 & 0 \\ 0 & 0 & \mathbf{r}_3 \end{pmatrix}, \; \mathcal{R}_\vee = \begin{pmatrix} \mathbf{\rho}_2 & 0 & 0 \\ 0 & \mathbf{\rho}_3 & 0 \\ 0 & 0 & \mathbf{R}_g \end{pmatrix}$$

The subscripts $\wedge$ and $\vee$ symbolize 'up-to-down' and 'down-to-up' reflections. Since $\mathcal{E}$ is known, (32) yields the light climate between medium layers in the form

$$J = \mathcal{G}\mathcal{E}, \qquad \mathcal{G} \equiv (1-\mathcal{Q})^{-1}, \qquad \mathbf{U}_0 = \mathbf{u}_1 + \mathbf{\rho}_1 \mathbf{D}_0 + \mathbf{\tau}_1 \mathbf{U}_1 \qquad (34)$$

The series $\mathcal{G} = 1 + \mathcal{Q} + \mathcal{Q}^2 + ...$ sums over all possible wanderings of photons among the $M$ medium layers, hence necessarily converges. Since $\mathcal{Q}$ is non-negative, so is $\mathcal{G}$.

**Block form of $\mathcal{G}$:** Note that $\mathcal{T}_d$, $\mathcal{T}_u$ are nilpotent, $(\mathcal{T}_v)^M = 0$, $v = d, u$, so that $(1-\mathcal{T}_v)^{-1} = 1 + \mathcal{T}_v + \mathcal{T}_v^2 + ... + \mathcal{T}_v^{M-1}$. Explicitly (for $M = 3$):

$$\mathbf{a}^{-1} \equiv (1-\mathcal{T}_d)^{-1} = \begin{pmatrix} 1 & 0 & 0 \\ \mathbf{t}_2 & 1 & 0 \\ \mathbf{t}_3\mathbf{t}_2 & \mathbf{t}_3 & 1 \end{pmatrix}, \qquad \mathbf{d}^{-1} \equiv (1-\mathcal{T}_u)^{-1} = \begin{pmatrix} 1 & \mathbf{\tau}_2 & \mathbf{\tau}_2\mathbf{\tau}_3 \\ 0 & 1 & \mathbf{\tau}_3 \\ 0 & 0 & 1 \end{pmatrix} \qquad (35)$$

Since $\mathcal{G}$ and $\mathbf{d}^{-1}$ exist, one can write $\mathcal{G}$ in the 'down-up' block form [11]:

$$\mathcal{G} = \frac{1}{1-\mathcal{Q}} = \begin{pmatrix} \mathcal{G}_{DD} & \mathcal{G}_{DU} \\ \mathcal{G}_{UD} & \mathcal{G}_{UU} \end{pmatrix}, \qquad 1-\mathcal{Q} = \begin{pmatrix} 1-\mathcal{T}_d & -\mathcal{R}_\wedge \\ -\mathcal{R}_\vee & 1-\mathcal{T}_u \end{pmatrix} \equiv \begin{pmatrix} \mathbf{a} & -\mathbf{b} \\ -\mathbf{c} & \mathbf{d} \end{pmatrix} \qquad (36)$$

$$\mathcal{G}_{DD} = \frac{1}{\mathbf{a} - \mathbf{b}\mathbf{d}^{-1}\mathbf{c}}, \quad \mathcal{G}_{UU} = \mathbf{d}^{-1} + \mathbf{d}^{-1}\mathbf{c}\mathcal{G}_{DD}\mathbf{b}\mathbf{d}^{-1}, \quad \mathcal{G}_{DU} = \mathcal{G}_{DD}\mathbf{b}\mathbf{d}^{-1}, \quad \mathcal{G}_{UD} = \mathbf{d}^{-1}\mathbf{c}\mathcal{G}_{DD} \qquad (37)$$

**Physical meaning of $\mathcal{G}$:** Observing that

$$\begin{aligned} J_{(m,j)} &= \sum_{m'',j''} \mathcal{G}_{(m,j),(m'',j'')} \mathcal{E}_{(m'',j'')} & (a) \\ J_{(m,j)} &= \mathcal{G}_{(m,j),(m',j')} \qquad \text{if} \quad \mathcal{E}_{(m'',j'')} = \delta_{m'm''}\delta_{j'j''} & (b) \end{aligned} \qquad (38)$$

we see that $J$ is a sum of subclimates due to each elementary 'emission' $\mathcal{E}_{(m'',j'')}$, while $\mathcal{G}_{(m,j),(m',j')}$ is the flux in direction $j$ at height $z_m$, due to a unit 'emission' in direction $j'$ by layer $m'$ (the usual meaning of a Green's matrix). Thus, another way to build $\mathcal{G}$ is as follows: Put all $\mathcal{E}_{(m,j)} = 0$, except for the first, $\mathcal{E}_{(1,1)} = 1$. Then the resulting light climate (computed by whatever means, for instance by iterating (30)) constitutes the first column of $\mathcal{G}$. Putting $\mathcal{E}_{(1,2)} = 1$, all others zero, yields the second column, and so forth.

## 9  Summary of TTRG

We may now summarize TTRG as follows:

(i) Partition the canopy into $N$ 'thin' layers $(z_{n-1}, z_n)$, such that $\mathbf{M}(z)$ and $\mathbf{E}(z)$ vary little within each 'thin' layer, and such that all $|\Delta z_n \mathbf{M}_{ij}(\bar{z}_n)| \ll 1$.

(ii) Compute the transfer matrices $\mathbf{T}_n \equiv \mathbf{T}(z_n, z_{n-1}) = e^{\mathbf{sM}_n \Delta z_n}$ (which differ little from $\mathbf{1}$).

(iii) Build $\mathbf{T}(z_n, z_0) = \mathbf{T}_n \mathbf{T}_{n-1} \cdots \mathbf{T}_1$ for $n = 1, 2, \ldots$, and check at each $n$ that no $|\mathbf{T}_{ij}(z_n, z_0)|$ exceeds some preset maximum $T_{\max}$ (say $10^8$ if double precision is used). If $T_{\max}$ is not exceeded after all the $\mathbf{T}_n$'s have been multiplied together, yielding the transfer matrix for the whole canopy, then one can proceed as in section 4. But in general:

(iv) If some $|\mathbf{T}_{ij}(z_n, z_0)|$ exceeds $T_{\max}$ at $n = n_1$ say, then call $(z_0, z_{n_1})$ the first 'medium' layer, with $\mathbf{T}^{(1)} \equiv \mathbf{T}(z_{n_1}, z_0)$. Then form $\mathbf{T}_{(n_1+l)} \mathbf{T}_{(n_1+l-1)} \cdots \mathbf{T}_{(n_1+1)}$, $l = 1, 2, \ldots$, checking again for large matrix elements. If $T_{\max}$ is again exceeded, at $n_1 + l = n_2$ say, then call $(z_{n_1}, z_{n_2})$ the second 'medium' layer, with $\mathbf{T}^{(2)} \equiv \mathbf{T}(z_{n_2}, z_{n_1})$. And so on until the whole canopy has been partitioned into $M$ 'medium' layers, with transfer matrices $\mathbf{T}^{(m)}$.

(v) If only the 'up' flux $\mathbf{U}_0$ at the top $z_0$ is required, compute $\mathbf{R}_0, \mathbf{e}_0$ by the 'discrete invariant embedding method' of section 7, and deduce $\mathbf{U}_0 = \mathbf{R}_0 \mathbf{D}_0 + \mathbf{e}_0$.

(vi) If $\mathbf{J}(z)$ at all heights $z$ are required, compute transmission-reflection matrices and emission vectors for the 'medium' layers, and construct $\mathcal{Q}$. Then compute the Green's matrix $\mathcal{G} = (1 - \mathcal{Q})^{-1}$, and apply it to the 'emission' vector $\mathcal{E}$ to get the fluxes between medium layers.

(vii) Finally, determine the fluxes between thin layers, *within* each medium layer, by using the transfer matrix equation (15) recursively.

**Part III  Other methods for numerically integrating the LTE**

This part describes other ways of computing light climates, for comparison with TTRG. We mostly discuss the widely used iterative integration method. That method produces the multiple-scattering expansion (or variants of it), which allows to estimate its rate of convergence. We also explain why a pure Green's matrix approach is not interesting, and discuss the invariant embedding method (which concerns only light reflected by the canopy).

**10  Pure Green's matrix approach**

Equations identical to (30) of course also hold for the fluxes $\mathbf{J}_n$ between the $N$ *thin* layers throughout the canopy. That is, putting $\mathbf{u}_{N+1} \equiv \mathbf{e}_g$, $\boldsymbol{\rho}_{N+1} \equiv \mathbf{R}_g$, $\boldsymbol{\tau}_{N+1} \equiv 0$:

$$\begin{aligned}
\mathbf{D}_n &= \mathbf{d}_n + \mathbf{t}_n \mathbf{D}_{n-1} + \mathbf{r}_n \mathbf{U}_n & (n = 1, 2, \ldots, N) & \quad (a) \\
\mathbf{U}_{n-1} &= \mathbf{u}_n + \boldsymbol{\rho}_n \mathbf{D}_{n-1} + \boldsymbol{\tau}_n \mathbf{U}_n & (n = 1, 2, \ldots N+1) & \quad (b)
\end{aligned} \quad (39)$$

Again, a Green's matrix $\mathcal{G}^{thin}$ may be used to solve (39). Once $\mathcal{G}^{thin}$ is known, it can be applied to any 'emission' vector $\mathcal{E}^{thin}$ to yield at once *all* the fluxes $\mathbf{J}_n$:

$$J^{thin}_{(n,j)} \equiv (\mathbf{J}_n)_j = \sum_{n',j'} \mathcal{G}^{thin}_{(n,j),(n',j')} \mathcal{E}^{thin}_{(n',j')} \quad (40)$$

A first drawback, however, is that $\mathcal{G}^{thin}$ is a $N_G^{thin} \times N_G^{thin}$ matrix, where $N_G^{thin} = N \times N_J$ is much larger than $N_G = M \times N_J$ in (34) (since $N \gg M$), which may stress machine capabilities. More importantly, using $\mathcal{G}^{thin}$ would be much more time-consuming than using the *medium* layer Green's matrix $\mathcal{G}$, followed by the recursion (15) inside each medium layer. Indeed, $J_{n,j}$ depends on *all* the elementary 'emissions' $\mathcal{E}^{thin}_{(n',j')}$ in (40), but only on $\mathbf{J}_{n-1}$ and $\mathbf{e}_n$ in $\mathbf{J}_n = \mathbf{T}_n \mathbf{J}_{n-1} + \mathbf{se}_n$, implying much fewer multiplications if used recursively (unless 'emission' occurs in only a few thin layers, e.g., if there is only incident light).

**11  Multiple-scattering expansion (MSE)**

An expansion of the light climate in numbers of scatterings suffered will now be written down. Viewing $\mathbf{D}_0$ and $\mathbf{e}_g$ as 'emissions', and $\mathbf{R}_g$ as reflection, from the 'edges' $z_0$ and $z_g$ of the canopy, we define a 'total' emission $\mathbf{E}^{tot}$, and 'total' scattering matrix $\mathbf{H}^{tot}$, by

$$\begin{aligned}
\mathbf{E}_D^{tot}(z) &\equiv \mathbf{E}_D(z) + \mathbf{D}_0 \delta(z - z_0), \qquad \mathbf{E}_U^{tot}(z) \equiv \mathbf{E}_U(z) + \mathbf{e}_g \delta(z - z_g) \\
\mathbf{H}_{UD}^{tot}(z) &\equiv \mathbf{H}_{UD}(z) + \mathbf{R}_g \delta(z - z_g)
\end{aligned} \quad (41)$$

*Unscattered light*: Recall that a photon travelling in direction $j$ has probability $G_j(z)dz$ to hit a leaf over an infinitesimal vertical distance $dz$. It follows that its probability $W_j^f(z,z')$ to go *freely* from height $z'$ to $z$ satisfies $(d/dz)W_j^f = -G_j W_j^f$, hence is given by

$$W_j^f(z,z') = e^{-\Lambda_j(z,z')}, \qquad \Lambda_j(z,z') \equiv \left| \int_{z'}^{z} dz'' G_j(z'') \right| \tag{42}$$

Thus, the *unscattered* light $\mathbf{J}^{(0)}(z)$, namely light that got directly to $z$ after being 'emitted', without suffering any scatterings by leaves or ground, is given by

$$\mathbf{D}^{(0)}(z) = \int_{z_0}^{z} dz' e^{-\Lambda^D(z,z')} \mathbf{E}_D^{tot}(z'), \qquad \mathbf{U}^{(0)}(z) = \int_{z}^{z_g} dz' e^{-\Lambda^U(z,z')} \mathbf{E}_U^{tot}(z') \tag{43}$$

*Multiple-scattering expansion*: We now express the light climate as a sum

$$\mathbf{J}(z) = \mathbf{J}^{(0)}(z) + \mathbf{J}^{(1)}(z) + \mathbf{J}^{(2)}(z) + \ldots \tag{44}$$

where $\mathbf{J}^{(k)}(z)$, $k = 1, 2, \ldots$, is light that got scattered $k$ times before reaching $z$. These are generated by the physically obvious recursions, for $k = 1, 2, \ldots$:

$$\begin{aligned}\mathbf{D}^{(k)}(z) &= \int_{z_0}^{z} dz' e^{-\Lambda^D(z,z')} \left[ \mathbf{H}_{DD}(z') \mathbf{D}^{(k-1)}(z') + \mathbf{H}_{DU}(z') \mathbf{U}^{(k-1)}(z') \right] \\ \mathbf{U}^{(k)}(z) &= \int_{z}^{z_g} dz' e^{-\Lambda^U(z,z')} \left[ \mathbf{H}_{UD}^{tot}(z') \mathbf{D}^{(k-1)}(z') + \mathbf{H}_{UU}(z') \mathbf{U}^{(k-1)}(z') \right] \end{aligned} \tag{45}$$

For instance, reading matrix products from right to left, $e^{-\Lambda^D(z,z')} \mathbf{H}_{DU}(z') \mathbf{U}^{(k-1)}(z')$ is $(k-1)$-times scattered 'up' light that gets scattered once more at $z'$, downwards, and then 'freely' propagates to $z$.

*Number of terms needed*: How many terms in (44) are needed for a given precision depends on how absorbing the leaves and ground are. If absorption is important (more than 80% for visible and thermal light), then about 3 iterations suffice to get $1\%$ precision relative to the unscattered light, since 3 scatterings reduce fluxes by $(0.2)^3 = 0.008$. With less absorption (about 20% for near infrared light), more terms are needed (about 20). Thus, in realistic situations, the numbers of terms needed are fairly small. However, in (artificial) extreme light trapping situations, these numbers can get very large (hundreds of thousands); moreover, just specifying a convergence criterium can become tricky.

## 12 Iterative integration

Iterative integration consists in iterating the *thin* layer equations (39) with $\mathbf{d}_n, \mathbf{u}_n$ and $\mathbf{t}_n, \mathbf{r}_n, \boldsymbol{\rho}_n, \boldsymbol{\tau}_n$ treated to first order in $\Delta z_n$. Of course, one could equally well iterate (39) as it stands, or iterate the 'medium' layer equations (30).

*Thin layer approximation*: To first order in $\Delta z$, one has in view of (6)-(7):

$$\mathbf{K} \approx 1 + \mathbf{m}, \qquad \boldsymbol{\varepsilon} \approx \mathbf{e}, \qquad \mathbf{m} \equiv \mathbf{h} - \mathbf{g} \tag{46}$$

so that $\mathbf{t} \approx 1 - \mathbf{g}_D + \mathbf{h}_{DD}$, $\mathbf{r} \approx \mathbf{h}_{DU}$, $\mathbf{d} \approx \mathbf{e}_D$, etc. These approximations in effect assume that along a light beam, the interception and scattering rates remain constant across the layer, whereas in reality they diminish with the attenuation of the beam. Thus, (46) suppose too much light blocked or reflected, making transmissions $\mathbf{t}, \boldsymbol{\tau}$ too small, reflections $\mathbf{r}, \boldsymbol{\rho}$ too large. The consequence will be, in general, less penetration of incident light into the canopy. The emissions $\mathbf{d}, \mathbf{u}$ can err either way, being increased by the neglect of attenuation on the way out, but decreased by the neglect of photons emitted in the opposite direction and then reflected back.

*Iteration*: Set all $\mathbf{U}_n \equiv \mathbf{u}_{n+1}$ in (39), and get values for $\mathbf{D}_n$ by using (a) for $n = 1, 2, ..., N$; then get new values for $\mathbf{U}_n$ by using (b) for $n = N+1, N, ..., 2$. Repeat using these new values of $\mathbf{U}_n$. And so on until values stabilize. Finally, get $\mathbf{U}_0$ using (b) with $n = 1$. This process generates a sequence $\mathbf{J}_n^{\{k\}}$, $k = 0, 1, 2, ...$, recursively: $\mathbf{D}_0^{\{k\}} \equiv \mathbf{D}_0$, $\mathbf{U}_n^{\{-1\}} \equiv \mathbf{u}_{n+1}$, and

$$\begin{aligned} \mathbf{D}_n^{\{k\}} &= \mathbf{d}_n + \mathbf{t}_n \mathbf{D}_{n-1}^{\{k\}} + \mathbf{r}_n \mathbf{U}_n^{\{k-1\}} & (n = 1, 2, ..., N) \\ \mathbf{U}_{n-1}^{\{k\}} &= \mathbf{u}_n + \boldsymbol{\tau}_n \mathbf{U}_n^{\{k\}} + \boldsymbol{\rho}_n \mathbf{D}_{n-1}^{\{k\}} & (n = N+1, N, ..., 1) \end{aligned} \tag{47}$$

*Stability*: Everything in (47) is non-negative. If $\mathbf{J}_n^{\{k\}}$ and $\mathbf{J}_n^{\{k-1\}}$ are equal to the exact $\mathbf{J}_n^{exact}$, then the terms added to $\mathbf{d}_n$ and $\mathbf{u}_n$ on the right of (47) are such as to yield again $\mathbf{D}_n^{exact}$ and $\mathbf{U}_{n-1}^{exact}$ on the left. So if some $\mathbf{J}_{n,j}^{\{k\}}$ and/or $\mathbf{J}_{n,j}^{\{k-1\}}$ are smaller than $\mathbf{J}_{n,j}^{exact}$, then the amounts added are less than should be, so that $\mathbf{J}_n^{\{k\}}$ on the left *remain* smaller than $\mathbf{J}_n^{exact}$. Yet $\mathbf{J}_n^{\{k\}}$ increase at each iteration. It follows, since initially $\mathbf{U}_n^{\{-1\}} \equiv \mathbf{u}_{n+1}$ are smaller than $\mathbf{U}_n^{exact}$, that $\mathbf{J}_n^{\{k\}}$ approaches $\mathbf{J}_n^{exact}$ *from below* as $k \to \infty$. Hence convergence is assured (as long as the $\mathbf{J}_n^{exact}$ are finite). To estimate how many iterations are needed for a given precision, we must analyse the physical meaning of $\mathbf{J}_n^{\{k\}}$. For this it is easier to first study a variant of (47).

## 13 A variant producing the multiple-scattering expansion [1]

Refering to (42) and (46), let us write $\mathbf{t}_n = \mathbf{t}_n^f + \mathbf{t}_n^s$, $\boldsymbol{\tau}_n = \boldsymbol{\tau}_n^f + \boldsymbol{\tau}_n^s$, where

$$\begin{aligned} \mathbf{t}_n^f &\equiv e^{-\Lambda^D(z_n, z_{n-1})} \approx 1 - \mathbf{g}_{nD}, & \boldsymbol{\tau}_n^f &\equiv e^{-\Lambda^U(z_{n-1}, z_n)} \approx 1 - \mathbf{g}_{nU} \\ \mathbf{t}_n^s &\equiv \mathbf{t}_n - \mathbf{t}_n^f \approx \mathbf{h}_{nDD}, & \boldsymbol{\tau}_n^s &\equiv \boldsymbol{\tau}_n - \boldsymbol{\tau}_n^f \approx \mathbf{h}_{nUU} \end{aligned} \tag{48}$$

so that $\mathbf{t}_n^f, \boldsymbol{\tau}_n^f$ give the probabilities for a photon to pass freely through 'thin' layer $n$, without hitting a leaf, and $\mathbf{t}_n^s, \mathbf{\tau}_n^s$ those to get scattered *onwards* by the layer, that is, transmitted after getting scattered any number ($\geq 1$) of times inside the layer. In the following, 'scattered' means 'scattered by a layer' in the above sense; it recovers its meaning 'scattered *once*' when $\mathbf{t}_n^s, \mathbf{\tau}_n^s, \mathbf{r}_n, \mathbf{\rho}_n$ are treated to first order in $\Delta z_n$.

***Iteration***: Define now a sequence $\mathbf{J}_n^{[k]}$, $k = 0,1,2,...$, by $\mathbf{D}_0^{[k]} \equiv \mathbf{D}_0$, and

$$\begin{aligned}
\mathbf{D}_n^{[k]} &= \mathbf{d}_n + \mathbf{t}_n^f \mathbf{D}_{n-1}^{[k]} + \mathbf{t}_n^s \mathbf{D}_{n-1}^{[k-1]} + \mathbf{r}_n \mathbf{U}_n^{[k-1]} & (n=0,1,2,...,N) \\
\mathbf{U}_{n-1}^{[k]} &= \mathbf{u}_n + \mathbf{\tau}_n^f \mathbf{U}_n^{[k]} + \mathbf{\tau}_n^s \mathbf{U}_n^{[k-1]} + \mathbf{\rho}_n \mathbf{D}_{n-1}^{[k-1]} & (n=N+1,N,...,1)
\end{aligned} \quad (49)$$

or, in an obvious notation following (29):

$$\mathbf{J}_n^{out,[k]} = \mathbf{\varepsilon}_n + \mathbf{K}_n^f \mathbf{J}_n^{in,[k]} + \mathbf{K}_n^s \mathbf{J}_n^{in,[k-1]}, \qquad \mathbf{D}_0^{[k]} \equiv \mathbf{D}_0 \qquad (50)$$

The zeroeth cycle ($k = 0$) is initialized by $\mathbf{D}_n^{[-1]} \equiv \mathbf{U}_n^{[-1]} \equiv 0$ for *all* $n$. The scattering matrices $\mathbf{K}_n^s$ then act on $\mathbf{J}_n^{in,[-1]} \equiv 0$, so that the zeroeth cycle yields the unscattered light $\mathbf{J}^{(0)}$. In the next cycle ($k=1$), $\mathbf{K}_n^s$ act on $\mathbf{J}^{[0]} = \mathbf{J}^{(0)}$, hence produce 'once-scattered' light $\mathbf{J}^{(1)}$. Thus $\mathbf{J}^{[1]} = \mathbf{J}^{(0)} + \mathbf{J}^{(1)}$. And, in general, $\mathbf{J}^{[k]} = \mathbf{J}^{(0)} + \mathbf{J}^{(1)} + ... + \mathbf{J}^{(k)}$. Explicitly, for $N = 3$:

$$\begin{aligned}
\mathbf{D}_1^{[0]} &= \mathbf{d}_1 + \mathbf{t}_1^f \mathbf{D}_0 = \mathbf{D}_1^{(0)}, \qquad \mathbf{D}_2^{[0]} = \mathbf{d}_2 + \mathbf{t}_2^f \mathbf{d}_1 + \mathbf{t}_2^f \mathbf{t}_1^f \mathbf{D}_0 = \mathbf{D}_2^{(0)} \\
\mathbf{D}_3^{[0]} &= \mathbf{d}_3 + \mathbf{t}_3^f \mathbf{d}_2 + \mathbf{t}_3^f \mathbf{t}_2^f \mathbf{d}_1 + \mathbf{t}_3^f \mathbf{t}_2^f \mathbf{t}_1^f \mathbf{D}_0 = \mathbf{D}_3^{(0)} \\
\mathbf{U}_3^{[0]} &= \mathbf{e}_g = \mathbf{U}_3^{(0)}, \quad \mathbf{U}_2^{[0]} = \mathbf{u}_3 + \mathbf{\tau}_3^f \mathbf{e}_g = \mathbf{U}_2^{(0)}, \quad \mathbf{U}_1^{[0]} = \mathbf{u}_2 + \mathbf{\tau}_2^f \mathbf{u}_3 + \mathbf{\tau}_2^f \mathbf{\tau}_3^f \mathbf{e}_g = \mathbf{U}_1^{(0)} \\
\mathbf{D}_1^{[1]} &= \left(\mathbf{d}_1 + \mathbf{t}_1^f \mathbf{D}_0\right) + \left(\mathbf{t}_1^s \mathbf{D}_0 + \mathbf{r}_1 (\mathbf{u}_2 + \mathbf{\tau}_2^f \mathbf{u}_3 + \mathbf{\tau}_2^f \mathbf{\tau}_3^f \mathbf{e}_g)\right) = \mathbf{D}_1^{(0)} + \mathbf{D}_1^{(1)}
\end{aligned} \quad (51)$$

and so forth. If $\mathbf{d}_n, \mathbf{t}_n$, etc., are treated to first order in $\Delta z_n$, then these equations actually build up the integrals (45) according to the mid-point numerical integration rule. The above iterative integration provides, in fact, the simplest way of generating the MSE (44).

***Number of iterations needed***: In (49)-(50), the scattering gains of $\mathbf{D}_n^{[k]}$ and $\mathbf{U}_{n-1}^{[k]}$ (the $\mathbf{K}^s$ terms) are computed using $\mathbf{D}_{n-1}^{[k-1]}$ and $\mathbf{U}_n^{[k-1]}$ *one cycle back*. In (47), by contrast, scattering gains use the 'newest' radiances available. Hence, $\mathbf{J}^{\{k\}}$ contains more terms than $\mathbf{J}^{[k]}$. For instance, $\mathbf{D}_1^{\{0\}} = \mathbf{d}_1 + \mathbf{t}_1 \mathbf{D}_0 + \mathbf{r}_1 \mathbf{u}_2 = \mathbf{D}_1^{(0)} + \left(\mathbf{t}_1^s \mathbf{D}_0 + \mathbf{r}_1 \mathbf{u}_2\right)$, where the second term is a piece of $\mathbf{D}_1^{(1)}$ (it misses $\mathbf{r}_1 \mathbf{\tau}_2^f \mathbf{u}_3 + \mathbf{r}_1 \mathbf{\tau}_2^f \mathbf{\tau}_3^f \mathbf{e}_g$ in (51)). In general, $\mathbf{J}^{\{k\}}$ equals $\mathbf{J}^{[k]}$ plus pieces of $\mathbf{J}^{(k+l)}$ for $k+l$ up to $2kN$. Yet, from the argument after (47), we know that $\mathbf{J}^{\{k\}}$ builds up the correct $\mathbf{J} = \mathbf{J}^{(0)} + \mathbf{J}^{(1)} + \mathbf{J}^{(2)} + ...$ as $k \to \infty$. Iteration of (47) or of (49) require the same number of multiplications per cycle, but in (47) each cycle incorporates more terms. Still, since $\mathbf{J}^{\{k\}}$ misses pieces of $\mathbf{J}^{(k+1)}$, the number of iterations of (47) needed for a given precision is roughly the same as for (49), that is, as the number of terms needed in the MSE (44). Hence, the discussion at the end of section 11 applies here as well.

## 14 Multiple-scattering expansion of the Green's matrix

The Green's matrix $\mathcal{G}$ in (34) is not adapted to getting the MSE. For this one may consider a different Green's matrix (which is only of theoretical interest). Rewrite equations (30) as

$$\tilde{\mathcal{J}} = \tilde{\mathcal{E}} + \tilde{\mathcal{Q}}\,\tilde{\mathcal{J}}, \qquad \tilde{\mathcal{J}} = \begin{pmatrix}\tilde{\mathcal{J}}_D \\ \tilde{\mathcal{J}}_U\end{pmatrix}, \qquad \tilde{\mathcal{E}} = \begin{pmatrix}\tilde{\mathcal{E}}_D \\ \tilde{\mathcal{E}}_U\end{pmatrix}, \qquad \tilde{\mathcal{Q}} = \begin{pmatrix}\tilde{\mathcal{T}}_d & \tilde{\mathcal{R}}_\wedge \\ \tilde{\mathcal{R}}_\vee & \tilde{\mathcal{T}}_u\end{pmatrix} \tag{52}$$

$$\tilde{\mathcal{J}}_D = \begin{pmatrix}\mathbf{D}_0 \\ \mathbf{D}_1 \\ \mathbf{D}_2 \\ \mathbf{D}_3\end{pmatrix}, \quad \tilde{\mathcal{J}}_U = \begin{pmatrix}\mathbf{U}_0 \\ \mathbf{U}_1 \\ \mathbf{U}_2 \\ \mathbf{U}_3\end{pmatrix}, \quad \tilde{\mathcal{E}}_D = \begin{pmatrix}\mathbf{D}_0 \\ \mathbf{d}_1 \\ \mathbf{d}_2 \\ \mathbf{d}_3\end{pmatrix}, \quad \tilde{\mathcal{E}}_U = \begin{pmatrix}\mathbf{u}_1 \\ \mathbf{u}_2 \\ \mathbf{u}_3 \\ \mathbf{e}_g\end{pmatrix}$$

$$\tilde{\mathcal{T}}_d = \begin{pmatrix}0 & 0 & 0 & 0 \\ \mathbf{t}_1 & 0 & 0 & 0 \\ 0 & \mathbf{t}_2 & 0 & 0 \\ 0 & 0 & \mathbf{t}_3 & 0\end{pmatrix}, \quad \tilde{\mathcal{T}}_u = \begin{pmatrix}0 & \mathbf{\tau}_1 & 0 & 0 \\ 0 & 0 & \mathbf{\tau}_2 & 0 \\ 0 & 0 & 0 & \mathbf{\tau}_3 \\ 0 & 0 & 0 & 0\end{pmatrix}, \quad \tilde{\mathcal{R}}_\wedge = \begin{pmatrix}0 & 0 & 0 & 0 \\ 0 & \mathbf{r}_1 & 0 & 0 \\ 0 & 0 & \mathbf{r}_2 & 0 \\ 0 & 0 & 0 & \mathbf{r}_3\end{pmatrix}, \quad \tilde{\mathcal{R}}_\vee = \begin{pmatrix}\mathbf{\rho}_1 & 0 & 0 & 0 \\ 0 & \mathbf{\rho}_2 & 0 & 0 \\ 0 & 0 & \mathbf{\rho}_3 & 0 \\ 0 & 0 & 0 & \mathbf{R}_g\end{pmatrix} \tag{53}$$

where $\tilde{\mathcal{J}}$ includes $\mathbf{D}_0$ and $\mathbf{U}_0$, unlike $\mathcal{J}$ in (33). Again, $\tilde{\mathcal{J}} = \tilde{\mathcal{G}}\tilde{\mathcal{E}}$ with $\tilde{\mathcal{G}} = (1-\tilde{\mathcal{Q}})^{-1}$. Using the breakup (48) of $\mathbf{t},\mathbf{\tau}$ into 'free' and 'scatter' parts, denote

$$\tilde{\mathcal{Q}} = \tilde{\mathcal{Q}}_0 + \tilde{\mathcal{V}}, \qquad \tilde{\mathcal{Q}}_0 = \begin{pmatrix}\tilde{\mathcal{T}}_d^f & 0 \\ 0 & \tilde{\mathcal{T}}_u^f\end{pmatrix}, \qquad \tilde{\mathcal{V}} = \begin{pmatrix}\tilde{\mathcal{T}}_d^s & \tilde{\mathcal{R}}_\wedge \\ \tilde{\mathcal{R}}_\vee & \tilde{\mathcal{T}}_u^s\end{pmatrix} \tag{54}$$

Denoting also $\tilde{\mathcal{G}}_0 = (1-\tilde{\mathcal{Q}}_0)^{-1}$, so that $\tilde{\mathcal{G}}_{0DD}$ and $\tilde{\mathcal{G}}_{0UU}$ are similar to (35), but with $\mathbf{t}^f, \mathbf{\tau}^f$ instead of $\mathbf{t},\mathbf{\tau}$, write [2]

$$\tilde{\mathcal{G}} \equiv (1-\tilde{\mathcal{Q}})^{-1} = \tilde{\mathcal{G}}_0(1-\tilde{\mathcal{V}}\tilde{\mathcal{G}}_0)^{-1} = \tilde{\mathcal{G}}_0 + \tilde{\mathcal{G}}_0\tilde{\mathcal{V}}\tilde{\mathcal{G}}_0 + \tilde{\mathcal{G}}_0(\tilde{\mathcal{V}}\tilde{\mathcal{G}}_0)^2 + \ldots \tag{55}$$

Acting this on $\tilde{\mathcal{E}}$ yields a multiple-scattering (by layers) expansion of the light climate $\tilde{\mathcal{J}}$. By contrast, $\tilde{\mathcal{Q}}^k\tilde{\mathcal{E}}$ is $k$-times scattered light but with 'traversing a layer without hitting a leaf' also considered 'scattering'. In (33), $\mathcal{Q}^k\mathcal{E}$ contains pieces of higher scattering orders since $\mathcal{E}$ contains $\mathbf{t}_1\mathbf{D}_0$. Thus, the various iterative integrations, and various Green's matrix series, all build up the light climate in different ways. The Green's matrix $\tilde{\mathcal{G}}$ is more 'physical', but $\mathcal{G}$ is better for computation since it is smaller.

---

[2] This is simply $(\mathbf{A}-\mathbf{B})^{-1} = \mathbf{A}^{-1}[\mathbf{1}-\mathbf{B}\mathbf{A}^{-1}]^{-1}$ with $\mathbf{A} = 1-\tilde{\mathcal{Q}}_0$, $\mathbf{B} = \tilde{\mathcal{V}}$. Alternatively, iterate $\tilde{\mathcal{G}} = \tilde{\mathcal{G}}_0 + \tilde{\mathcal{G}}_0\tilde{\mathcal{V}}\tilde{\mathcal{G}}$, following from $(\mathbf{A}-\mathbf{B})^{-1} = \mathbf{A}^{-1} + \mathbf{A}^{-1}\mathbf{B}(\mathbf{A}-\mathbf{B})^{-1}$.

## 15 Composition rules for $\mathbf{K}, \boldsymbol{\varepsilon}$

Consider two layers, $a = (z_1, z_2)$ (above), and $b = (z_2, z_3)$ (below), with $z_1 < z_2 < z_3$. Applying (21) to each layer yields:

$$\begin{pmatrix} \mathbf{D}_2 \\ \mathbf{U}_1 \end{pmatrix} = \begin{pmatrix} \mathbf{t}_a & \mathbf{r}_a \\ \boldsymbol{\rho}_a & \boldsymbol{\tau}_a \end{pmatrix} \begin{pmatrix} \mathbf{D}_1 \\ \mathbf{U}_2 \end{pmatrix} + \begin{pmatrix} \mathbf{d}_a \\ \mathbf{u}_a \end{pmatrix}, \qquad \begin{pmatrix} \mathbf{D}_3 \\ \mathbf{U}_2 \end{pmatrix} = \begin{pmatrix} \mathbf{t}_b & \mathbf{r}_b \\ \boldsymbol{\rho}_b & \boldsymbol{\tau}_b \end{pmatrix} \begin{pmatrix} \mathbf{D}_2 \\ \mathbf{U}_3 \end{pmatrix} + \begin{pmatrix} \mathbf{d}_b \\ \mathbf{u}_b \end{pmatrix} \tag{56}$$

Let now $\mathbf{K}, \boldsymbol{\varepsilon}$ pertain to the combined pair of layers, namely $(z_1, z_3)$, so that

$$\mathbf{J}^{out} \equiv \begin{pmatrix} \mathbf{D}_3 \\ \mathbf{U}_1 \end{pmatrix} = \begin{pmatrix} \mathbf{t} & \mathbf{r} \\ \boldsymbol{\rho} & \boldsymbol{\tau} \end{pmatrix} \begin{pmatrix} \mathbf{D}_1 \\ \mathbf{U}_3 \end{pmatrix} + \begin{pmatrix} \mathbf{d} \\ \mathbf{u} \end{pmatrix} = \mathbf{K} \mathbf{J}^{in} + \boldsymbol{\varepsilon}, \qquad \mathbf{J}^{in} = \begin{pmatrix} \mathbf{D}^{in} \\ \mathbf{U}^{in} \end{pmatrix} \equiv \begin{pmatrix} \mathbf{D}_1 \\ \mathbf{U}_3 \end{pmatrix} \tag{57}$$

Using (56), one easily obtains the 'composition rules' (that for $\mathbf{K}$ is well known [7]):

$$\mathbf{K} = \begin{pmatrix} \mathbf{t}_b \mathcal{R}_{ab} \mathbf{t}_a & \mathbf{r}_b + \mathbf{t}_b \mathcal{R}_{ab} \mathbf{r}_a \boldsymbol{\tau}_b \\ \boldsymbol{\rho}_a + \boldsymbol{\tau}_a \mathcal{R}_{ba} \boldsymbol{\rho}_b \mathbf{t}_a & \boldsymbol{\tau}_a \mathcal{R}_{ba} \boldsymbol{\tau}_b \end{pmatrix}, \qquad \boldsymbol{\varepsilon} = \begin{pmatrix} \mathbf{d}_b + \mathbf{t}_b \mathcal{R}_{ab} (\mathbf{d}_a + \mathbf{r}_a \mathbf{u}_b) \\ \mathbf{u}_a + \boldsymbol{\tau}_a \mathcal{R}_{ba} (\mathbf{u}_b + \boldsymbol{\rho}_b \mathbf{d}_a) \end{pmatrix} \tag{58}$$

$$\mathcal{R}_{ab} \equiv \frac{1}{1 - \mathbf{r}_a \boldsymbol{\rho}_b}, \qquad \mathcal{R}_{ba} \equiv \frac{1}{1 - \boldsymbol{\rho}_b \mathbf{r}_a} = 1 + \boldsymbol{\rho}_b \mathcal{R}_{ab} \mathbf{r}_a, \qquad \boldsymbol{\rho}_b \mathcal{R}_{ab} = \mathcal{R}_{ba} \boldsymbol{\rho}_b \tag{59}$$

The physical meanings are clear: For instance (reading matrix products from right to left), the 'down' transmission $\mathbf{t} = \mathbf{t}_b \mathcal{R}_{ab} \mathbf{t}_a$ is transmission $\mathbf{t}_a$ into the interlayer, followed by multiple reflections $\mathcal{R}_{ab} = 1 + \mathbf{r}_a \boldsymbol{\rho}_b + \mathbf{r}_a \boldsymbol{\rho}_b \mathbf{r}_a \boldsymbol{\rho}_b + ...$ between the two layers, and finally transmission $\mathbf{t}_b$ to the outside. The radiances between the two layers, $(\mathbf{D}^\mu, \mathbf{U}^\mu) \equiv (\mathbf{D}_2, \mathbf{U}_2)$, are

$$\begin{aligned} \mathbf{D}^\mu &= \mathcal{R}_{ab} \left( \mathbf{t}_a \mathbf{D}^{in} + \mathbf{r}_a \boldsymbol{\tau}_b \mathbf{U}^{in} + \mathbf{d}_a + \mathbf{r}_a \mathbf{u}_b \right) \\ \mathbf{U}^\mu &= \mathcal{R}_{ba} \left( \boldsymbol{\rho}_b \mathbf{t}_a \mathbf{D}^{in} + \boldsymbol{\tau}_b \mathbf{U}^{in} + \boldsymbol{\rho}_b \mathbf{d}_a + \mathbf{u}_b \right) \end{aligned} \tag{60}$$

***Relation to*** $\mathbf{R}_0, \mathbf{e}_0$: Observe that

$$\boldsymbol{\rho} = \boldsymbol{\rho}_a + \boldsymbol{\tau}_a \frac{1}{1 - \boldsymbol{\rho}_b \mathbf{r}_a} \boldsymbol{\rho}_b \mathbf{t}_a, \qquad \mathbf{u} = \mathbf{u}_a + \boldsymbol{\tau}_a \frac{1}{1 - \boldsymbol{\rho}_b \mathbf{r}_a} (\mathbf{u}_b + \boldsymbol{\rho}_b \mathbf{d}_a) \tag{61}$$

involve only reflection $\boldsymbol{\rho}_b$ and emission $\mathbf{u}_b$ from the *top* of layer $b$. Thus, if we let $a$ be the canopy, and the top surface of $b$ play the role of the ground, then $\boldsymbol{\rho} = \mathbf{R}_0$, $\mathbf{u} = \mathbf{e}_0$, $\boldsymbol{\rho}_b = \mathbf{R}_g$, $\mathbf{u}_b = \mathbf{e}_g$, and (61) is equivalent to (20) (via (22)-(23)).

***Light climate***: The radiances between 'medium' layers may now be obtained as follows: Let layer $\overline{m}$ consist of all the canopy *below* layer $m$, and compute $\mathbf{K}_{\overline{m}}, \boldsymbol{\varepsilon}_{\overline{m}}$ recursively by using (58) with $a = m+1$, $b = \overline{m+1}$, for $m = M-1, M-2,...,0$. The last step yields $\mathbf{K}_{\overline{0}} = \mathbf{K}$, $\boldsymbol{\varepsilon}_{\overline{0}} = \boldsymbol{\varepsilon}$ for the whole canopy. Then, $\mathbf{D}_g = \mathbf{t}\mathbf{D}_0 + \mathbf{r}\mathbf{U}_g + \mathbf{d}_g = \mathbf{t}\mathbf{D}_0 + \mathbf{r}(\mathbf{R}_g \mathbf{D}_g + \mathbf{e}_g) + \mathbf{d}_g$, whence

$$\mathbf{D}_g = \frac{1}{1-\mathbf{rR}_g}(\mathbf{t}\mathbf{D}_0 + \mathbf{re}_g + \mathbf{d}_g), \qquad \mathbf{U}_g = \mathbf{R}_g \mathbf{D}_g + \mathbf{e}_g \tag{62}$$

Knowing $\mathbf{D}_0$ and $\mathbf{U}_g$, deduce $\mathbf{J}_1$ by using (60) for layers 1 and $\overline{1}$. Then, knowing $\mathbf{D}_1$ and $\mathbf{U}_g$, get $\mathbf{J}_2$ using (60) for 2 and $\overline{2}$, and so on.

## 16 Differential equations for $\mathbf{K}, \boldsymbol{\varepsilon}$, invariant embedding method

By setting $z_2 = z_0$, $z_3 = z_g$ and $z_2 - z_1 = dz_0$ in (58), so that $\mathbf{K}_a \approx \mathbf{1} + \mathbf{M}(z_0)dz_0$, $\boldsymbol{\varepsilon}_a \approx \mathbf{E}(z_0)dz_0$, by (8), one obtains, denoting $\dot{g} \equiv -dg/dz_0$:[3]

$$\dot{\mathbf{K}} = \begin{pmatrix} \dot{\mathbf{t}} & \dot{\mathbf{r}} \\ \dot{\boldsymbol{\rho}} & \dot{\boldsymbol{\tau}} \end{pmatrix} = \begin{pmatrix} \mathbf{tA} + \mathbf{tB}\boldsymbol{\rho} & \mathbf{tB}\boldsymbol{\tau} \\ \boldsymbol{\rho}\mathbf{A} + \boldsymbol{\rho}\mathbf{B}\boldsymbol{\rho} + \mathbf{C} + \mathbf{D}\boldsymbol{\rho} & \boldsymbol{\rho}\mathbf{B}\boldsymbol{\tau} + \mathbf{D}\boldsymbol{\tau} \end{pmatrix}, \qquad \mathbf{M}(z_0) \equiv \begin{pmatrix} \mathbf{A} & \mathbf{B} \\ \mathbf{C} & \mathbf{D} \end{pmatrix}$$

$$\dot{\boldsymbol{\varepsilon}} = \begin{pmatrix} \dot{\mathbf{d}} \\ \dot{\mathbf{u}} \end{pmatrix} = \begin{pmatrix} \mathbf{tBu} + \mathbf{tE}_D \\ \boldsymbol{\rho}\mathbf{Bu} + \mathbf{Du} + \boldsymbol{\rho}\mathbf{E}_D + \mathbf{E}_U \end{pmatrix}, \qquad \mathbf{E}(z_0) \equiv \begin{pmatrix} \mathbf{E}_D \\ \mathbf{E}_U \end{pmatrix} \tag{63}$$

By rather setting $z_1 = z_0$, $z_2 = z_g$, $z_3 - z_2 = dz_g$ in (58), one finds that $\mathbf{K}' \equiv d\mathbf{K}/dz_g$, $\boldsymbol{\varepsilon}' \equiv d\boldsymbol{\varepsilon}/dz_g$ are given by (63) with 'up' and 'down' interchanged, and $z_g$ instead of $z_0$.

*Invariant embedding method*: Observe from (63) that

$$\dot{\boldsymbol{\rho}} = \boldsymbol{\rho}\mathbf{A} + \boldsymbol{\rho}\mathbf{B}\boldsymbol{\rho} + \mathbf{C} + \mathbf{D}\boldsymbol{\rho}, \qquad \dot{\mathbf{u}} = \boldsymbol{\rho}\mathbf{Bu} + \mathbf{Du} + \boldsymbol{\rho}\mathbf{E}_D + \mathbf{E}_U \tag{64}$$

involve only $\boldsymbol{\rho}, \mathbf{u}$ and $\mathbf{M}(z_0), \mathbf{E}(z_0)$. Indeed, how an additional layer $dz_0$ affects reflection and emission from the top depends only on *local* quantities at the top, not on what goes on below. Thus, (64) also hold in the presence of a reflecting and emitting ground below, in which case $\boldsymbol{\rho} = \mathbf{R}_0$ and $\mathbf{u} = \mathbf{e}_0$, see (20). The equation for $\dot{\boldsymbol{\rho}} = \dot{\mathbf{R}}_0$ is a matrix *Ricatti* equation. Its numerical integration is quite stable, and is the basis of the so-called *invariant embedding method* [10]. The multistep calculation of $\mathbf{R}_0, \mathbf{e}_0$ in section 7, in effect using (61), is clearly an accelerated 'large step' version of that procedure.

*Milne problem* [9]: In a very thick (realistic) canopy, light penetrates only so far, so that the rest of the canopy below is irrelevant. One can evaluate $\rho = R_0$, $u = e_0$ by integrating $K', \varepsilon'$ from $z_0$ downwards, or more quickly by adding finite layers *below* using (58), until $\rho, u$ stabilize. In the case of a *uniform* canopy, one can rather add layers *above*, starting at $z_0$ with $R_0 = 0$, $e_0 = 0$, and use the simpler equations (61) or (64), which do not involve $r, \tau, d$.

---

[3] Alternatively, use (22) and $\dot{\mathbf{T}} = \mathbf{TsM}$, $\dot{\mathbf{f}} = \mathbf{TsE}$, by (11) and (14). For instance, for $\boldsymbol{\tau} = \mathbf{T}_{UU}^{-1}$ one gets: $\dot{\boldsymbol{\tau}} = -\mathbf{T}_{UU}^{-1}\dot{\mathbf{T}}_{UU}\mathbf{T}_{UU}^{-1} = -\mathbf{T}_{UU}^{-1}(\mathbf{T}_{UD}\mathbf{M}_{DU} - \mathbf{T}_{UU}\mathbf{M}_{UU})\mathbf{T}_{UU}^{-1} = \boldsymbol{\rho}\mathbf{M}_{DU}\boldsymbol{\tau} + \mathbf{M}_{UU}\boldsymbol{\tau}$.

## 17 Conclusion

We presented a method (TTRG) for numerically integrating the light transport equation, in a canopy above a reflecting-emitting ground. Other methods were reviewed for comparison. The focus was on the widely used iterative integration method (ITINT) (which produces the multiple-scattering expansion). ITINT (to first order in $\Delta z_n$) has the advantage of requiring only the matrices $G_n, H_n$, but at the cost of a loss in precision. The latter can be improved by making $\Delta z_n$ smaller, which, however, increases computation times. Alternatively, one can include higher order corrections, for instance use [4] $K = 1 + m + \frac{1}{2}m^2$, $\varepsilon = (1 + \frac{1}{2}m)e$ to second order in $\Delta z$ (compare (46)). But then one may as well use the exact $K_n, \varepsilon_n$. Obviously, it is faster to use TTRG, which requires $K_m, \varepsilon_m$ for only a few 'medium' layers, and then $\mathcal{G}$. All these matrices need to be computed only once for a given canopy, and then applied to various 'emission' vectors to rapidly yields light climates.[5]

We performed a number of numerical tests regarding the speed and accuracy of TTRG, as compared to ITINT. These tests will be described in detail elsewhere. But we here give the main points.

We work in double precision (i.e., carrying 15 digits). The number of discrete photon directions is either $N_J = 18$ (zenithal inclinations $\theta_j$ only, as is usual), or $N_J = 18 \times 18 = 324$ directions $(\theta_j, \varphi_j)$. Canopies have a leaf area index $LAI = 10$ (total leaf area per unit horizontal area of canopy). The propagated emissions $f_n$ are computed using Simpson's rule (the transfer matrix equations (15) are modified accordingly). Note that iterative integration amounts to the

---

[4] These are best obtained by using linear fractional maps, as will be described elsewhere.

[5] An alternative to using the Green's matrix $\mathcal{G}$, to get the radiances between 'medium' layers, is the method described at the end of section 15, using the composition rules for $K, \varepsilon$.

midpoint or trapezoid rule.[6]

In order to assess the precisions of numerical computations, we considered situations admitting *analytical solutions*, specifically: Canopies uniform in $z$, with azymuthally isotropic leaf area densities of non-absorbing (but emitting) Lambertian leaves, with different coefficients for the top and bottom sides of leaves (allowing to create extreme light trapping situations). In these cases, the matrices $\mathbf{M}$ are known analytically, permitting high precision tests.

When there are no leaf emissions, precision with TTRG is better than $10^{-10}$. With leaf emissions, overall precision is limited by that of the Simpson's integration done to get $\mathbf{f}_n$, namely between $10^{-4}$ and $10^{-3}$ with 'thin' layers of $\text{LAI} \approx 0.1$ (better with thinner layers). Such high precisions cannot be contemplated with ITINT (especially with grazing sunlight). With ITINT, precision is less than 10% if thin layers have $\text{LAI} \approx 0.1$, and 0.5% with thin layer $\text{LAI} \approx 0.01$. In general, ITINT reduces penetration of incident light into the canopy, because thin layer transmissions $\mathbf{t}_n, \boldsymbol{\tau}_n$ are too small, and reflections $\mathbf{r}_n, \boldsymbol{\rho}_n$ too large, when treated to first order in $\Delta z_n$ (as explained in section 12).

Concerning speed, let us first discuss TTRG applied to a realistic canopy of $\text{LAI} = 10$, using $N_J = 324$ angular sectors. We use $N = 100$ thin layers of $\text{LAI} = 0.1$. We only mention those quantities which take a significant time to compute.[7] Each thin layer transfer matrix $\mathbf{T}_n = e^{\mathbf{m}_n}$ takes ~1.6 seconds, for a total of 160 secs if all the thin layers are different (in practice, field data separates the canopy into about 10 homogeneous layers, so that only 10 different $\mathbf{T}_n$ need to be computed, hence ~16 secs). Computing $\mathbf{T}^{(m)}$ for 5 'medium' layers of $\text{LAI} = 2$ (by multiplying thin layers $\mathbf{T}_n$ together) takes ~16 secs; getting $\mathbf{K}_m$ and $\mathcal{G}$ takes another 15 secs. Once all these matrices are known, getting a light climate takes ~0.5 secs.

ITINT does not require all these matrices. However, getting a light climate using 100 thin layers of $\text{LAI} \approx 0.1$ takes ~2 secs for visible light (high absorption), 11 secs for NIR (low absorption), but precision is less than 10% as said above. If 1000 thin layers of $\text{LAI} \approx 0.01$ are used, then precision is better than 0.5%, but computations times are 44 secs for visible, 221 secs

---

[6] According as one puts $\mathbf{e}_n = \mathbf{E}(\bar{z}_n)\Delta z_n$ or $\mathbf{e}_n = \frac{1}{2}\big(\mathbf{E}(z_n) + \mathbf{E}(z_{n+1})\big)\Delta z_n$ in (46).

[7] The computation times quoted are for a 2.66 GHz Pentium IV, with 1 GB of RAM. The matrix exponentiations and inversions are done in MATLAB.

for NIR.

With $N_J = 18$ angular sectors, TTRG takes a fraction of a second (including computation of all matrices). ITINT takes comparable times for realistic canopies. But in artificial situations of extreme light trapping, ITINT may take on the order of hours, for a precision of 1%. Also, specifying a convergence criterium becomes tricky. But TTRG remains as fast and precise.

Thus, for realistic canopies and $N_J = 18$ angular sectors, computation time by either method is not a significant issue if only a few light climates are needed. But ITINT is not too reliable (unless one checks convergence as $\Delta z_n \to 0$, which takes time). Time becomes an issue if the number of angular sectors $N_J$ is large (since the number of multiplications scales roughly as $N_J^2$). In any case, TTRG is much more accurate, and has a definite speed advantage whenever many light climates must be computed for the *same* canopy (as when iterating over leaf temperatures).

**Acknowledgments**: We thank A. Cernusca and U. Tappeiner for suggesting the problem and, together with G. Wohlfahrt, for useful discussions.